\documentclass{acm_proc_article-sp}
\usepackage{hyperref}
\usepackage{multirow}

\begin{document}

\title{Bermuda: Bidirectional de novo assembly of transcripts with new insights for handling uneven coverage }
%
%
%
%
%

\numberofauthors{5} 
%
\author{
%
%
\alignauthor
Qingming Tang\\
       \affaddr{Toyota Technological Institute at Chicago}\\
       \affaddr{6045 S. Kenwood Ave.}\\
       \affaddr{Chicago, Illinois 60637}\\
       \email{qmtang@ttic.edu}
\alignauthor
Sheng Wang\\
       \affaddr{Toyota Technological Institute at Chicago}\\
       \affaddr{6045 S. Kenwood Ave.}\\
       \affaddr{Chicago, Illinois 60637}\\
       \email{shengwang@ttic.edu}
\alignauthor Jian Peng\\
       \affaddr{University of Illinois at Urbana-Champaign}\\
       \affaddr{2118 Siebel Center}\\
       \affaddr{201 N Goodwin Ave Urbana, IL, 61801}\\
       \email{jianpeng@illinois.edu}
\and  
\alignauthor Jianzhu Ma\\
       \affaddr{Toyota Technological Institute at Chicago}\\
       \affaddr{6045 S. Kenwood Ave.}\\
       \affaddr{Chicago, Illinois 60637}\\
       \email{majianzhu@ttic.edu}
\alignauthor Jinbo Xu \footnotemark[1]\\
       \affaddr{Toyota Technological Institute at Chicago}\\
       \affaddr{6045 S. Kenwood Ave.}\\
       \affaddr{Chicago, Illinois 60637}\\
       \email{jinbo.xu@gmail.com}
}

\maketitle
\begin{abstract}
\textbf{Motivation}: RNA-seq has made feasible the analysis of a whole set of expressed mRNAs. Mapping-based assembly of RNA-seq reads sometimes is infeasible due to lack of high-quality references. However, de novo assembly is very challenging due to uneven expression levels among transcripts and also the read coverage variation within a single transcript. Existing methods either apply de Bruijn graphs of single-sized k-mers to assemble the full set of transcripts, or conduct multiple runs of assembly, but still apply graphs of single-sized k-mers at each run. However, a single k-mer size is not suitable for all the regions of the transcripts with varied coverage.\\
\textbf{Contribution}: This paper presents a de novo assembler Bermuda with new insights for handling uneven coverage. Opposed to existing methods that use a single k-mer size for all the transcripts in each run of assembly, Bermuda self-adaptively uses a few k-mer sizes to assemble different regions of a single transcript according to their local coverage. As such, Bermuda can deal with uneven expression levels and coverage not only among transcripts, but also within a single transcript. 
Extensive tests show that Bermuda outperforms popular de novo assemblers in reconstructing unevenly-expressed transcripts with longer length, better contiguity and lower redundancy. Further, Bermuda is computationally efficient with moderate memory consumption.\\

\end{abstract}

\footnotetext[1]{Correspondence should be addressed to Dr. Xu}

\section{Introduction}
The advent of RNA-seq technique has revolutionized the study of transcriptomics as it promises a comprehensive transcriptional landscape. Compared to microarrays, RNA-seq can be used to annotate novel genes \cite{roberts2011identification, trapnell2010transcript}, refine the 5$'$ and 3$'$ ends of genes \cite{graveley2008haplo}, identify the boundary of exons and introns and analyze differential gene and transcript expression \cite{jiang2009statistical, rapaport2013comprehensive, trapnell2012differential, xia2011nsmap}. Further, RNA-seq can be applied to construct the whole set of expressed transcripts, which is one of the computational challenges in the field \cite{garber2011computational, martin2011next, nagarajan2013sequence}. Several reference-based assemblers (i.e., mapping assemblers) \cite{garber2011computational, guttman2010ab, trapnell2012differential, yassour2009ab} are developed to reconstruct the full set of expressed mRNAs. With high-quality references, even some erroneous reads can be correctly mapped and thus, expressed transcripts can be accurately reconstructed. \\
However, a high-quality reference genome is not always availa-ble. In fact we do not have reference genomes for the majority of the organisms. Further, due to the genomic difference among the sub-populations within one organism \cite{bryc2010genome, weir1984estimating}, the sample under assembly may be quite different from its reference at some regions. That is, sometimes we cannot rely on the reference even if it does exist. Therefore, de novo assemblers, which do not rely on references, are needed for transcript assembly. A few de novo assemblers \cite{birol2009novo, chu2013ebardenovo, grabherr2011full, koren2012hybrid, peng2013idba, robertson2010novo, schulz2012oases, surget2010optimization, xie2014soapdenovo} have been developed, but they still underperform mapping assemblers (e.g., Cufflinks \cite{trapnell2012differential}) when a good reference is available. 
\subsection{Challenges of de novo transcript assembly}
Although significant advance has been made to de novo genome assembly \cite{ariyaratne2011pe, butler2008allpaths, nagarajan2013sequence, zerbino2008velvet}, de novo transcriptome assembly is quite different from genome assembly and faces the following challenges.
\begin{figure*}[!Ht]
\centering
\includegraphics[width=0.9\textwidth]{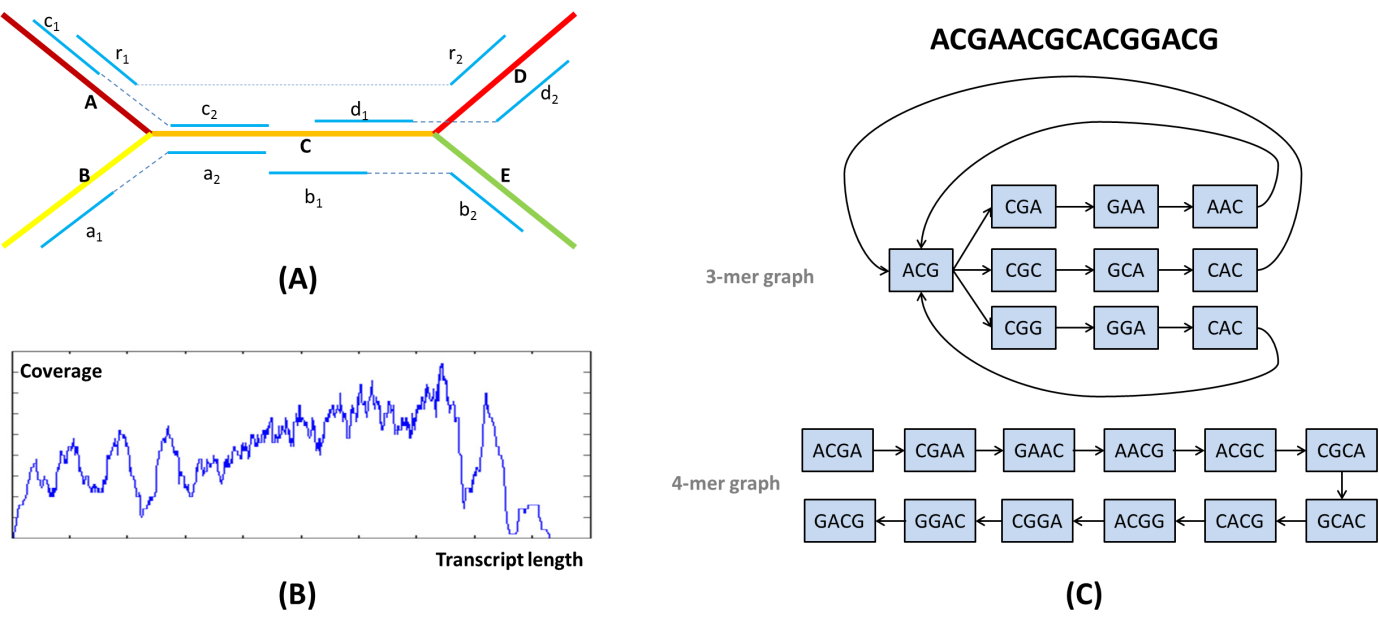}
\caption{Illustration of challenges of de novo transcript assembly. (A) isoform deconvolution. Five exons (A, B, C, D and E) form a splicing graph and [$r_1$(gap)$r_2$], [$a_1$(gap)$a_2$], [$b_1$(gap)$b_2$], [$c_1$(gap)$c_2$] and [$d_1$(gap)$d_2$] are all paired-end short reads. (B) Read coverage ($y$-axis) at each position ($x$-axis) of exon AT3G12580-E.2. Only those reads of SRR391052 that can align to this exon with at least 140 matched bases are considered. (C) Impact of $k$-mer size to de Bruijn graph.}
\label{Intro_Fig_1}
\end{figure*}
\subsubsection{Isoform identification}
Isoforms pose challenge to both de novo and mapping based assembly \cite{heber2002splicing, xing2004multiassembly}. Due to alternative splicing in the genome of eukaryotes, one gene may be spliced into multiple transcripts (i.e., isoforms) sharing exons. Paired-end sequencing, which has been applied in the study of genome sequences \cite{ariyaratne2011pe, campbell2008identification, chaisson2009novo, korbel2007paired}, can help resolve isoforms by making use of sequential order (or even sequential distance) of two short reads in the same molecule. In Figure \ref{Intro_Fig_1}A, the read [$r_1$(gap)$r_2$] clearly conveys the information that exon \textbf{A} is a prefix of exon \textbf{D} in some transcript. However, since the insert size (i.e., genomic distance between the $5'$ and $3'$ sequences of one paired-end read) is usually limited, merely utilizing paired-end information may not resolve isoforms. For example, still shown in Figure 1A, \textbf{ACD} and \textbf{BCE} are two isoforms. Suppose that two paired-end reads [$a_1$(gap)$a_2$] and [$b_1$(gap)$b_2$] are sequenced from \textbf{ACD} and two paired-end reads [$c_1$(gap)$c_2$] and [$d_1$(gap)$d_2$] are from \textbf{BCE} respectively. Only based upon these four reads we cannot infer that '\textbf{A}' must connect to '\textbf{D}' and '\textbf{B}' to '\textbf{E}'. Instead, '\textbf{ACE}' and '\textbf{BCD}' may also be feasible transcripts. 
\subsubsection{Uneven expression levels}
RNA-seq data has unevenly distributed coverage on transcripts under assembly due to a large range of expression. This leads to:
\begin{itemize}
\item It is challenging to select a good $k$-mer size for a de Bruijn graph assembler to conduct assembly. Highly-expressed transcripts or isoform-shared exons tend to have higher coverage while lowly-expressed transcripts lower coverage. De Bruijn graphs of a single $k$-mer size may not be suitable for the reconstruction of transcripts with varied expression levels. The reads for the transcripts with low expression tend to have less overlap and thus, these transcripts are less likely to be represented as a connected path in a de Bruijn graph of a large $k$-mer size. However, a smaller $k$ may reduce the assembly specificity. As shown in Figure \ref{Intro_Fig_1}C, there is a one-to-one correspondence between the sequence \textbf{ACGAACGCACGGACG} and the $4$-mer de Bruijn graph, but this is not true for the $3$-mer de Bruijn graph.
\item It is very challenging to correct sequencing errors. Ideally, the full set of expressed transcripts can be represented as many small disconnected de Bruijn graphs \cite{peng2013idba}, each connected graph may correspond to isoforms of one gene. However, due to the existence of erroneous $k$-mer nodes, these disconnected components may be coupled together, and traversing such a graph with chimeric edges may generate chimeric transcripts. Most de Bruijn graph genome assemblers, such as Velvet \cite{zerbino2008velvet} and ALLPATHS \cite{butler2008allpaths}, correct erroneous nodes by simply removing those $k$-mers with low support since they are very likely to be erroneous. However, such a method is not suitable for RNA-seq due to uneven expression levels. The support of an erroneous $k$-mer from a highly-expressed transcript may be higher than that of a correct $k$-mer from a lowly-expressed transcript. Therefore, simply removing $k$-mers with weak support will not only remove errors, but also many correct $k$-mers extracted from lowly-expressed transcripts.
\end{itemize}

\subsubsection{Coverage variation within a single transcript}
To better handling the uneven coverage on transcripts, some transcriptome assemblers conduct multiple runs of assembly, and apply a single but different $k$-mer de Bruijn graph at each run \cite{birol2009novo, peng2013idba, schulz2012oases}. However, this strategy cannot deal well with coverage variation in a single transcript. Coverage variation not only exists among transcripts, but also appears within even a single transcript. As shown in Figure \ref{Intro_Fig_1}B, the coverage of the exon AT3G12580-E.2 shows big variation. Actually, the sequencing process can be understood as a random process \cite{dohm2008substantial, mcintyre2011rna}, and each base pair can be hit by the random sequencing process with a certain probability in one run. Thus, for those transcripts with not very high expression levels, the variance of coverage would probably be more significant.

\subsection{Related work}
Several de novo transcript assemblers have been developed, including the Trinity \cite{grabherr2011full}, Oases-M \cite{schulz2012oases}, Trans-ABySS \cite{robertson2010novo}, IDBA-Tran \cite{peng2013idba} and SOAPdenovo-Trans \cite{xie2014soapdenovo}. All these assemblers are de Bruijn graph assemblers.\\
Trinity is developed to assemble highly-expressed transcripts with large support. Trinity uses a simple clustering-based error correction method and thus, may remove many correct $k$-mers sequenced from the transcripts with low expression. Further, Trinity uses only one de Bruijn graph of single and large-sized $k$-mer and thus, may miss the transcripts with low expression.\\
SOAPdenovo-Trans is another de Bruijn graph based assembler derived from the SOAPdenovo2 assembler \cite{luo2012soapdenovo2}, with an effective scaffolding module for transcript assembly. SOAP denovo-Trans utilizes the Trinity error correction strategy, a carefully designed de Bruijn graph simplification method and the Oases graph traversal strategy to generate high quality transcript candidates.\\
Unlike Trinity and SOAPdenovo-Trans, both Trans-ABySS and Oases-M use multiple de Bruijn graphs of different-sized $k$-mers to conduct assembly in multiple runs. In each run only one graph is used and the assembly results of all the runs are combined by a consensus method. Since they still use a single graph to construct each transcript fragment, they may not deal well with a transcript with large coverage variation across different regions \cite{dohm2008substantial, mcintyre2011rna}. Further, after combining the reconstructed transcripts from all the runs, they may produce duplicated (but different) copies of the same transcript and thus, generate many more transcripts than expected.\\
Similar to Oases-M, IDBA-Tran also uses multiple de Bruijn graphs with distinct $k$ to handle transcripts with different expression levels. IDBA-Tran employs a probabilistic progressive approach to iteratively remove erroneous $k$-mers with local thresholds. In each run, IDBA-Tran deals with the output of last round rather than the raw reads, which is different from Oases-M and Trans-ABySS, but IDBA-Tran still uses single $k$-mer de Bruijn graph to handle a group of transcripts.\\
All the methods described above can be regarded as "transcript-level" methods. That is, in each run of assembly, a single $k$-mer size is applied globally to all the transcripts.  

\subsection{Contribution}
In this study, we present a new de novo transcript assembler Ber-muda with the following features:
\begin{itemize}
\item Bermuda uses dynamic $k$-mer sizes to assemble different regions of a single transcript fragment in a single run. Consequently, it can deal with not only uneven expression levels among transcripts, but also coverage bias along a single transcript resulting from sequencing at random positions. By contrast, existing methods use single strategy (e.g. a single $k$-mer graph) to assemble the whole set of transcripts in one run. To the best of our knowledge, Bermuda is the first assembler that employs dynamic strategies for transcript assembly in a single run.
\item Bermuda uses de Bruijn graphs for only pre-assembly error correction and then directly joins overlapping short reads into transcripts, as opposed to existing methods that search connected paths in the graphs as the assembly results. By doing so, we can significantly reduce false positives (e.g., chimeric paths) introduced by de Bruijn graphs.
\item Bermuda uses a bi-directional assembly process to join overlapping short reads, which can significantly reduce the impact of insert size.
\end{itemize}

\section{MAERIALS AND METHODS}

\begin{figure}[!Ht]
\centering
\includegraphics[width=0.5\textwidth]{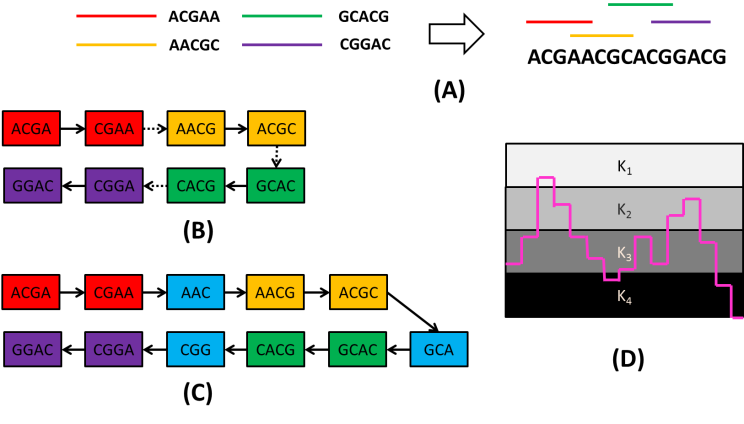}
\caption{Illustration of applying multiple $k$-mer sizes to assemble a single transcript. (A) The reads. (B) The corresponding $4$-mer de Bruijn graph built from reads. (C) The $3$,$4$-mer de Bruijn graph. (D) Applying four $k$ $(k_1>k_2>k_3>k_4)$ to assemble a transcript with varied coverage.}
\label{Intuition}
\end{figure}

\begin{figure}[!Ht]
\centering
\includegraphics[width=0.45\textwidth]{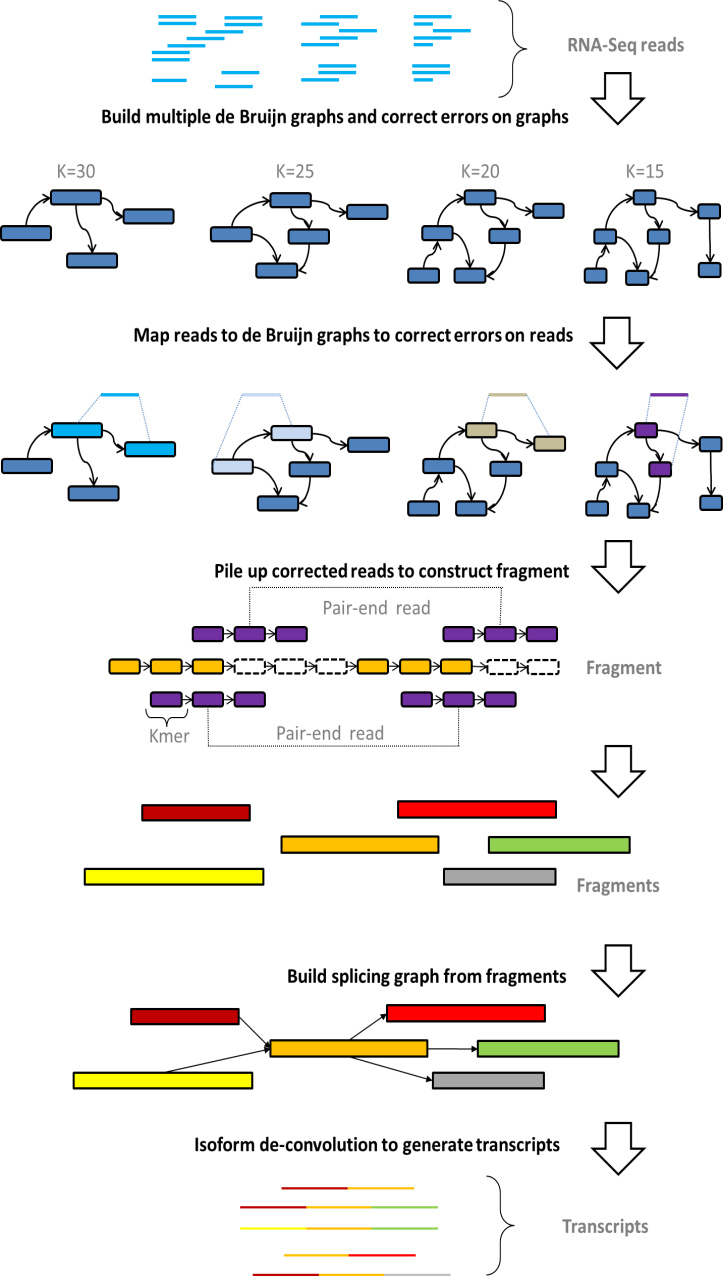}
\caption{Workflow of Bermuda.}
\label{Work_Flow}
\end{figure}

\subsection{New insights}
Existing de Bruijn graph assemblers either apply a de Bruijn graph of a carefully-chosen $k$-mer size to assemble the full set of transcripts, or conduct multiple runs of assembly and use a single-sized $k$-mer graph in each run. However, in some situations, it is not easy to find such a single $k$ so that high-quality assembly can be conducted using that $k$-mer graph. \\
As shown in Figure \ref{Intuition}A, supposing there are four reads for the sequence \textbf{ACGAACGCACGGACG}. Although the $3$-mer graph built from these reads is connected, it contains a few loops (see Figure \ref{Intro_Fig_1}C), which makes it hard to identify the true sequence among so many possible paths without extra information. If $4$-mers are used, as shown in Figure \ref{Intuition}B, it will result in four disconnected small de Bruijn graphs (marked by four different colors). That is, there is no single $k$ suitable for this case. However, if we use both $3$-mers and $4$-mers, then we can build a simple, linear and connected graph from the four reads, as shown in Figure \ref{Intuition}C.\\
The intuition is that we shall choose a $k$-mer size considering the read coverage at each position \cite{bankevich2012spades}. In particular, given a sequence $\{a_1,a_2,...,a_N\}$ and $M$ reads, the optimal $k$ for $a_i$ shall be the maximum number $s$, such that there exists a path $[a_{i-s},...,a_{i-1}] \to [a_{i-s+1},...,a_i ] \to [a_{i-s+2},...,a_{i+1}]$ in the $s$-mer de Bruijn graph. The larger read coverage at one point, it is more likely we can use a larger $k$-mer size.\\
Due to read coverage variation, it is infeasible to choose a specific $k$-mer size optimal for all positions. Instead, we may choose several distinct $k$-mer sizes, and adaptively choose one based on the local read coverage. In particular, we may divide the whole coverage range into four levels and at each level, we use a specific $k$, as shown in Figure \ref{Intuition}D. Generally speaking, larger $k$ is used for regions with higher coverage and smaller $k$ for regions with low coverage. 

\subsection{Overview of Bermuda}
Figure \ref{Work_Flow} illustrates the overall flowchart of Bermuda. Bermuda first corrects errors in reads, then it directly joins reads to build fragments. Afterwards, Bermuda links these fragments to form splicing graphs, each of which may correspond to isoforms of one or several genes, and finally traverses the graphs to generate candidate transcripts. To correct errors in reads, Bermuda first builds de Bruijn graphs, and conduct graph reduction in each graph. Then Bermuda maps reads back to the reduced de Bruijn graphs to correct sequencing errors in the reads. By default, Bermuda builds four types of de Bruijn graphs of $30-$, $25-$, $20-$, and $15-$mers, respectively. After read correction, Bermuda bi-directionally constructs transcript fragments one by one. For each transcript fragment under construction, Bermuda adaptively uses graphs of different $k$-mer sizes to join overlapping paired-end reads until the whole fragment has been constructed or no compatible reads are available. Bi-direction construction helps with low expression regions of a transcript and also isoform detection. Bermuda repeats the above construction procedure for each possible transcript fragment and also builds (alternative) splicing graphs for isoform detection. A splicing graph encodes all expressed isoforms of one gene or all transcripts of some similar genes. Finally, Bermuda employs the graph traversal algorithm of Oases \cite{schulz2012oases} to identify transcripts.

\begin{figure*}[!Ht]
\centering
\includegraphics[width=0.9\textwidth]{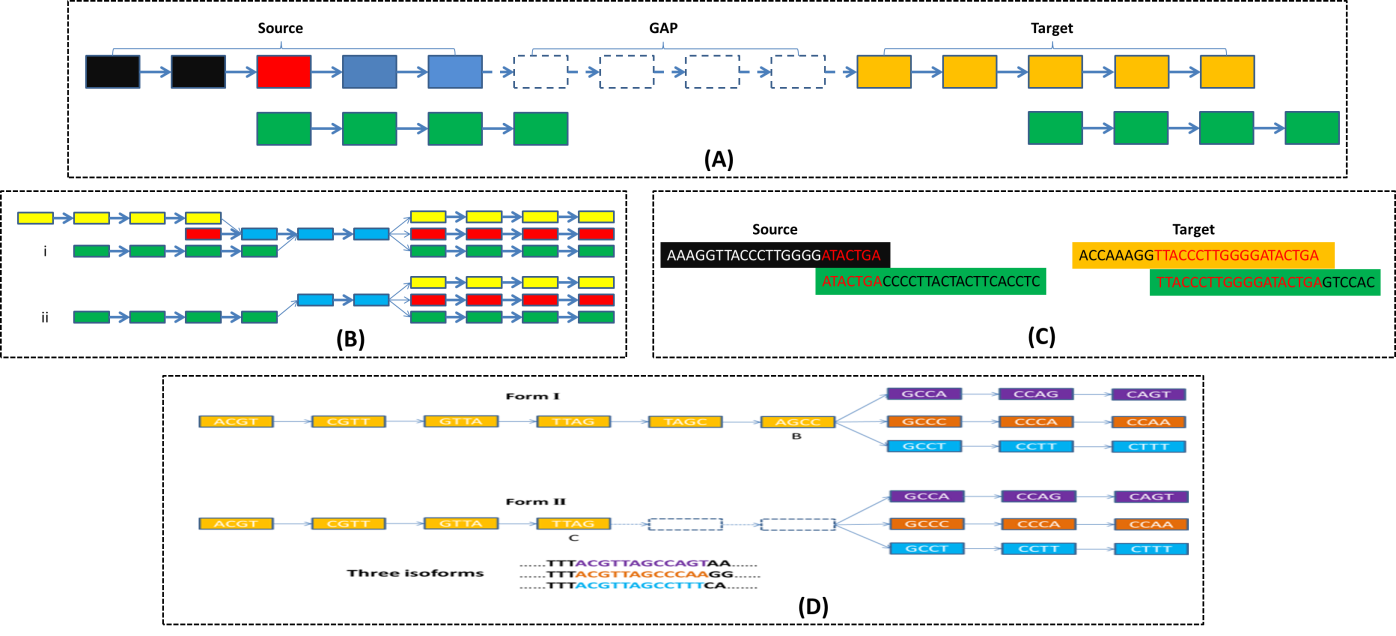}
\caption{Illustration of transcript fragment construction and splicing graph building. (A) Illustration of fragment growth process ($5'$ to $3'$). (B) Forks tracking (C) An example of reverse direction construction. (D) Illustration of two forms of forks. In Form $I$, node $B$ directly connects to three branches of the fork. In Form $II$, a gap exists between node $C$ and the three branches.}
\label{Construct}
\end{figure*}

\subsection{De Bruijn graphs construction and reduction}
Bermuda uses four types of $k$-mer de Bruijn graphs where $k$ equals to $15$, $20$, $25$, and $30$. A $k$-mer graph is represented as a set of $(k+1)$-mers, each representing an edge in the $k$-mer graph. For example, an edge \textbf{ACG} $\to$ \textbf{CGT} in a $3$-mer graph can be represented by a $4$-mer \textbf{ACGT}. The graph edges are stored in hash tables and balanced trees. See Supplementary Method for more technical details. \\
After building the de Bruijn graphs, Bermuda starts the graph reduction process to remove erroneous nodes in each of the de Bruijn graphs. Typically, a graph reduction process includes removing short tips and parallel paths that share the same start and end nodes with a true path \cite{kelley2010quake, koren2012hybrid, medvedev2011error, pevzner2001eulerian, robertson2010novo, schulz2012oases}. \\
However, a typical parallel path search algorithm may detect an erroneous path that sharing only its start and end nodes with a true path. It would be time-consuming to detect an erroneous path that shares its internal nodes with other true paths. Such kind of parallel paths may be more easily detected in a de Bruijn graph of larger $k$-mers as the internal nodes in an erroneous path may no longer match with nodes in other true paths when $k$ increases. Also, it would be impossible to correct errors in an isolated short path. However, in de Bruijn graph of small $k$-mers, such kind of isolated path may become either a short tip or an erroneous path parallel to some true path if the isolated short path is erroneous. So, by simultaneously examining multi-$k$-mer graphs, we can better remove erroneous nodes. For each single graph, we use a heuristic method to efficiently define the parallel paths. See Supplementary Method for more technical details about our parallel path searching algorithm for a de Bruijn graph.

\subsection{Read error correction}
We correct errors in the $5'$ and $3'$ sequences of one raw paired-end read separately, but with the same procedure. We borrow the ideas from existing error correction methods \cite{le2013probabilistic,schroder2009shrec,ilie2011hitec} but use multiple $k$-mers. To correct errors in a $5'$ or $3'$ sequence, we use the reduced $k$-mer de Bruijn graphs in the decreasing order of $k$ ($k=30$, $25$, $20$, and $15$) and execute the following two steps.
\begin{itemize}
\item If the first $k$-mer of the $5'$ sequence appears in the reduced $k$-mer graph, in the graph find the most similar path starting with this $k$-mer node for the sequence. Replace the $5'$ sequence by the path if they differ from each other by $1$ or $2$ positions. Otherwise, go to the next step.
\item Randomly select one $k$-mer (other than the first one) from the $5'$ sequence and use it as seed to find the most similar path for the sequence. Replace the $5'$ sequence by the path if they differ from each other by $1$ or $2$ bases. Otherwise, use reduced graph with smaller $k$-mer size and goes to step one.
\end{itemize}

\subsection{Transcript fragment construction}
Here we use "transcript fragment" to denote the concatenation of some compatible reads. A fragment may correspond to a single transcript, or multiple isoforms. 
\subsubsection{Basic fragment construction procedure}
We construct transcript fragments in both directions: from $5'$ to $3'$ and from $3'$ to $5'$. We grow a fragment by gradually adding and aligning compatible reads, using paired-end information and the context of both the $5'$ and $3'$ sequences. In the $5'$ to $3'$ direction, we say that one read is compatible with one node in the fragment under construction if the $5'$ sequence of this read starts from this node. Similarly, in the $3'$ to $5'$ direction, a read is compatible with one node if its $3'$ sequence starts from this node.\\
Since fragment construction in both directions is quite similar, we only describe the construction procedure from $5'$ to $3'$, as shown in Figure \ref{Construct}A. A fragment under construction consists of two parts: Source and Target. The Source and Target are initially formed by a pair of $5'$ and $3'$ sequences, respectively. They are disconnected and may become connected as the construction goes on. Each node in the Source has one of the three colors: black, red and blue, indicating the status of the reads compatible with the node. A black, red and blue node indicates that the compatible reads have been examined for the fragment growth, are currently under examination and are yet to be examined, respectively. At any time there is at most one red node. When current red node is done, we move to the blue node right next to the red and set its color to red if blue nodes exist.
As shown in Figure \ref{Construct}A, if the $5'$ and $3'$ sequences of one read compatible with the red node can be exactly aligned to the Source and Target respectively, the fragment can be linearly extended. By "exact alignment" we mean that the prefix of one sequence is exactly same as the suffix of the Source or Target. Otherwise, alternative paths (i.e., branches) are introduced to the Source or/and the Target, which results in a nonlinear structure called fork. The branches in a fork may stand for a true path or erroneous/chimeric paths. We merge a pair of branches if they differ by at most $2$ bases. Assume $x$ is the branch with the largest support, and let $|x|$ denote its support. Then any other branch with supports at least $0.05\times |x|$ is treated as a true path. Otherwise, it is treated as an erroneous path and accordingly removed. A fork with true branches encodes isoforms or paralogous genes. \\
Note that all the new nodes added to the Source are marked as blue. After examining all the nodes in the Source, we will also examine the nodes in the Target and continue fragment construction. 

\subsubsection{Using multiple k-mer size}
To handle transcripts with uneven expression levels or coverage bias along a single transcript, Bermuda makes use of four types of $k$-mer sizes, say $30-$, $25-$, $20-$, and $15-$mers, respectively and by default. Bermuda always chooses larger $k$-mers when there is sufficient coverage. Otherwise, it uses a smaller $k$-mers to increase sensitivity. More specifically, let $K$ denotes the size of the $k$-mers currently used. Whenever Bermuda finishes examining one red node (see Figure \ref{Construct}A), supposing that there are $x$ blue nodes right after the red node, Bermuda switches to the $k_1$-mer de Bruijn graph where $k_1$ is the largest integer (among $15$, $20$, $25$, and $30$) smaller than $K+x-1$. When $K=15$, and $x=0$, our construction algorithm terminates because we believe that overlapping smaller than 15 base pairs are not reliable. In such kind of case, fragment construction in reverse direction may be applied to continue the fragment construction as described in the following section.

\subsubsection{Fragment construction in reverse direction}
Fragment growth in reverse direction can help detect isoforms since fragment construction along a single direction may not be able to fully capture the complex structure of isoforms. For example, as in Figure \ref{Construct}B(i), we have a de Bruijn graph representation of three transcripts, which share a blue path in the middle. If we only use $5'$ to $3'$ construction starting from the green path, then we can only build a transcript fragment shown in Figure 4B(ii), which misses the red and yellow paths in the $5'$ side (see Figure \ref{Construct}B(i)). Nevertheless, if we conduct the construction along both directions, we may build the three transcripts shown in Figure \ref{Construct}B(i).
Transcript construction in reverse direction is also useful when the Source and Target of one fragment under construction is still disconnected after the $5'$ to $3'$ construction is done. As shown in Figure \ref{Construct}C, assume that there is one read (in green) overlapping with both the Source and Target. Due to the large variation of the insert size, the $5'$ sequence of this read overlaps with the Source by only $6$ bases (the $6$ bases overlap can be identified by brute force search or applying suffix tree), which is not a strong evidence indicating that the $5'$ sequence shall be joined to the Source. By contrast, the $3'$ sequence overlaps with the Target by many more bases. Therefore, if we extend the fragment from the Target to Source, the Source and Target may be connected. That is, reverse direction construction may help connect the Source to the Target or reduce the gap size between them. 

\subsection{Building splicing graphs}

As mentioned before, we use splicing graphs for isoform de-convolution. We build splicing graphs after constructing one transcript fragment in both directions ($5'$ to $3'$ and $3'$ to $5'$). When there is no fork in the constructed transcript fragment, there are no isoforms and hence, it is easy to deal with. 
If there are branches in either the Source or Target, there may be two types of forks, as shown in Figure \ref{Construct}D. When there is a common junction node connecting the branches as shown in Form I, we split the fragment at this junction node. The boundary of one fragment is defined by junction node. As in Figure \ref{Construct}D, node $B$ is the junction node for Form I and it defines the boundary for the sequence marked by yellow. When there is a gap (the blank nodes in Figure \ref{Construct}D) as shown in Form II of Figure \ref{Construct}D, we do not have a junction node as in Form I. As such, we find the nearest node (e.g., node $C$ in Form II) such that the reads passing through it would support at least two branches. We treat the nearest node as the junction node. Form II appears due to lack of reads to support the two blank nodes as shown in the graph. Finally, we split the frag-ment at all the junction nodes, treat each resultant linear segment as a vertex of the splicing graph and add one directed edge (from $5'$ to $3'$) between two segments that share a junction node and may belong to the same transcript. 

\section{RESULTS}
\begin{table*}[!Ht]
\caption{Statistics of assembly results by five assemblers.}
\centering
\begin{tabular}{|c|c|c|c|c|c|c|c|}
\hline
& $\#$Recovered & $\#$Candidates & $\#$Corrects & Pre1($\%$) & Aligned length & Unaligned length & Pre2 \\
\hline
\multicolumn{8}{|c|}{SRR364830 (human)}\\
\hline
Bermuda	& $10171$ & $89032$	& $34233$	& $36.4$	& $101445435$	& $28502762$ & $3.56$ \\
Trinity	& $8330$ & $120974$	& $34484$	& $28.5$	& $104505218$	& $33668196$ & $3.56$ \\
Oases	&-	&-	&-	&-	&-	&-	&- \\
IDBA	& $9194$	& $87483$	& $26898$	& $30.7$	& $93398029$	& $26190201$	& $3.57$ \\
SOAP	& $7017$	& $65562$	& $37902$	& $57.8$	& $59375913$	& $24522156$	& $2.42$ \\
\hline
\multicolumn{8}{|c|}{SRR391051 (thaliana)}\\
\hline
Bermuda	& $5752$ & $38043$	& $18071$	& $47.5$ &	$27854983$	& $15119906$ & $1.84$ \\
Trinity	& $4658$	& $48128$	& $20692$	& $43.0$ &	$22169248$	& $13177044$ &	$1.68$ \\
Oases	& $3642$ &	$112009$ &	$62435$	& $55.7$	& $33887243$	& $14028743$	& $2.42$ \\
IDBA	& $5439$	& $34121$	& $15893$	& $46.6$	& $23212486$	& $12599930$	& $1.84$ \\
SOAP	& $4123$	& $60918$	& $21528$	& $35.3$	& $20793459$	& $14103059$	& $1.47$ \\
\hline
\multicolumn{8}{|c|}{SRR404355 (mouse)}\\
\hline
Bermuda	& $1943$ &	$18876$	& $13207$	& $69.9$	& $20032435$	& $3132349$	& $6.40$ \\
Trinity	& $973$	& $24570$	& $16746$	& $68.2$	& $13514258$	& $1986079$	& $10.09$ \\
Oases	& $1277$ &	$80359$	& $57983$	& $71.1$	& $25825409$	& $2443274$	& $10.57$ \\
IDBA	& $1788$ &	$17733$	& $10984$	& $61.9$	& $17046935$	& $3253578$	& $5.24$ \\
SOAP	& $1443$	& $35051$ & $16529$ & 	$47.2$	& $15978392$	& $4930091$	& $3.24$ \\
\hline
\end{tabular}
\label{table:stat}
\end{table*}
\subsection{Datasets}
We benchmark our algorithm using the Human, Mouse and Arabidopsis Thaliana data sets. In total we have tested our method on $12$ datasets. However, due to space limit, here we present experimental results on three real data sets: SRR364830 (human), SRR391051 (thaliana) and SRR404355 (mouse). We compare our method Bermuda with four popular de novo transcript assemblers, including Trinity, Oases-M, IDBA-Tran and SOAPdenovo-Trans. To save space, we denote Oases-M as Oases, IDBA-Tran as IDBA, and SOAPdenovo-Trans as SOAP. All these assemblers use default or recommended parameter settings. We ran Oases using the supplied python pipeline (oases$\_$pipeline.py) with $k_{min}=23$ and $k_{max} = 31$. Such a pipeline will conduct a postprocessing to reduce redundancy in the results. For Bermuda, we use the $30$, $25$, $20$ and $15$-mer for error correction and fragment construction.
We pre-process all the raw reads by removing the reads containing little information evaluated as follows: for a continuous segment of 'N's, we assign a score of 20 to the first 'N' and a score of $2$ to each of the remaining 'N's. We assign a score of $1$ to each base other than 'N'. Let $P(N)$, $P(A)$, $P(C)$, $P(G)$ and $P(T)$ denote the total scores over 'N', 'A', 'C', 'G' and 'T' respectively, and $S$ be $P(A)+P(C)+P(G)+P(T)+P(N)$. We calculate the information content of one sequence as follows:
\begin{align}
&\frac{P(A)}{S}\ln{\frac{S}{P(A)}}+\frac{P(C)}{S}\ln{\frac{S}{P(C)}}\nonumber \\
&+\frac{P(G)}{S}\ln{\frac{S}{P(G)}}+\frac{P(T)}{S}\ln{\frac{S}{P(T)}}+\frac{P(N)}{S}\ln{\frac{S}{P(N)}}
\end{align}
We remove a paired-end read if either its $5'$ or $3'$ sequence has information content less than $0.5$. Some assemblers use the insert size information to assist transcript assembly, but we do not do so because the insert size may vary a lot.
\begin{table}[ht]
\caption{Performance (i.e., $\#$Recovered) with respect to expression levels.}
\centering
\begin{tabular}{|c|c|c|c|c|c|}
\hline
& [0-20] & [20-40] & [40-60] & [60-80] & [80-100] \\
\hline
\multicolumn{6}{|c|}{SRR364830 (human)}\\
\hline
Bermuda	& 632	& 2462	& 4637	& 4455	& 1032 \\
Trinity	& 417	& 1847	& 3571	& 3948	& 1019 \\
Oases	& -	& -	& -	& -	& - \\
IDBA	& 524	& 2155	& 4001	& 4263	& 952 \\
SOAP	& 350	& 1536	& 3047	& 3531	& 886 \\
\hline
\multicolumn{6}{|c|}{SRR391051 (thaliana)}\\
\hline
Bermuda	& 15	& 102	& 1403	& 3112	& 3478 \\
Trinity	& 2	& 51	& 710	& 2422	& 3211 \\
Oases	& 3	& 60	& 730	& 2317	& 3102 \\
IDBA	& 3	& 92	& 1228	& 3052	& 3428 \\
SOAP	& 1	& 42	& 794	& 2613	& 2897 \\
\hline
\multicolumn{6}{|c|}{SRR404355 (mouse)}\\
\hline
Bermuda	& 6	& 61	& 261	& 912	& 1723 \\
Trinity	& 1	& 9	& 60	& 362	& 1191 \\
Oases	& 1	& 17	& 92	& 491	& 1528 \\
IDBA	& 1	& 33	& 234	& 879	& 1693 \\
SOAP	& 0	& 26	& 215	& 828	& 1545 \\
\hline
\end{tabular}
\label{table:Performance_EXP}
\end{table}
\subsection{Experiment design and performance metrics }
To evaluate the predicted transcripts, we use BLAST \cite{altschul1997gapped} to align them to the corresponding cDNA databases. Human and mouse cDNA databases are downloaded from \hyperref{url}{http://useast.ensembl.org/index.html}, and Arabidopsis Thaliana cDNA database is from \hyperref{url}{http://plants.ensembl.org/index.html}. Only the best BLAST match for a predicted transcript is used to evaluate the performance.\\ 
We consider both completeness of a real transcript and correctness of a constructed candidate transcript. The completeness of one constructed candidate transcript is defined as the number of matched bases normalized by the length of the most similar cDNA. The completeness of one cDNA is the maximum completeness value of all the related alignments. The correctness of one constructed candidate is defined as the number of matched bases normalized by its length.\\
To further evaluate the performance of Bermuda, we analyze the constructed candidate transcripts at a given completeness level. Let $A$ denotes the set of candidates with completeness level larger than a given threshold. For each element $x\in A$, denote the best match of $x$ as $M(x)$. For one real cDNA $t$, if there are $r$ elements ${a_1,a_2,...,a_r}$ in $A$, such that $M(a_i)=t$ for $1\leq i \leq r$, then the redundancy of $t$ with respect to $A$ is $r-1$. Then the redundancy of $A$ is calculated by summing over the redundancy of all the cDNAs in the database.\\
To examine the performance of Bermuda at different expression levels, we align each RNA-Seq read set to the corresponding gene database using the Burrows-Wheeler Alignment tool \cite{li2009fast} and calculate the coverage of each gene as described in \cite{jiang2009statistical}. We divide all the expression levels into $5$ quantiles: $[0,20\%]$, $[20,40\%]$, $[40,60\%]$, $[60, 80\%]$, and $[80, 100\%]$, and calculate the number of recovered real transcripts in each quantile.
\begin{table}[ht]
\caption{Computational resources.}
\centering
\begin{tabular}{c c c c c}
\hline
\multicolumn{5}{c}{SRR364830 (human)}\\
\hline
& \multicolumn{3}{c}{Time (seconds)} & Memory \\
& 1 core & 2 cores & 4 cores & GB \\
\hline
Bermuda	& 35982	& 21337	& 11023	& 20.3 \\
Oases	& >86400	& >86400	& >86400	& > 126 \\
IDBA	& 107520	& 60323	& 32234	& 16.3 \\
SOAP	& 32880	& 17141	& 8823	& 24 \\
\hline
\multicolumn{5}{c}{SRR391051 (thaliana)}\\
\hline
& \multicolumn{3}{c}{Time (seconds)} & Memory \\
& 1 core & 2 cores & 4 cores & GB \\
\hline
Bermuda	& 9157	& 6231	& 4075	& 14.2 \\
Oases	& 140400	& 80565	& 49978	& 57.5 \\
IDBA	& 22380	& 14640	& 7680	& 8.5 \\
SOAP	& 10175	& 5973	& 3998	& 8.2 \\
\hline
\multicolumn{5}{c}{SRR404355 (mouse)}\\
\hline
& \multicolumn{3}{c}{Time (seconds)} & Memory \\
& 1 core & 2 cores & 4 cores & GB \\
\hline
Bermuda	& 1395	& 832	& 510 & 0.5 \\
Oases	& 1875	& 1043	& 605	& 
30.9 \\
IDBA	& 6135	& 3857	& 2713	& 2.1 \\
SOAP	& 1081	& 669	& 491	&7.0 \\
\hline
\end{tabular}
\label{table:Resources}
\end{table}
\subsection{Statistics of assembly results}
Table 1 lists the statistic information on the output of the five assemblers. Meanwhile, "$\#$Recovered" is the number of real transcripts that can be covered by some candidates at $\geq 90\%$ completeness level. Since the cDNA databases are fixed, so "$\#$Recovered" can be interpreted as the recall. "$\#$Candidates" is the number of the constructed candidate transcripts and "$\#$Corrects" is the number of candidates with $95\%$ correctness. The contig level precision (Pre$1$) is defined as "$\#$Corrects" normalized by "$\#$Candidates". "Aligned length" is the total number of bases in the predicted transcripts that can be aligned to the cDNA database using BLAST and "Unaligned length" is the number of bases that cannot be aligned. The nucleotide level precision (Pre2) is defined as aligned length normalized by unaligned length. Oases failed to run on the SRR364830 (human) data set because it consumes more than $126G$ RAM, which is unavailable on all the computers to which we can access. \\
As shown in Table 1, Bermuda outperforms all the other assemblers in terms of the number of reconstructed transcripts (i.e., "$\#$Recovered"). Oases has the best performance in terms of the contig and nucleotide level precision, but very low recall. This may be because Oases generates many short contigs, which can be easily aligned to cDNA databases very well and thus increase both the contig and nucleotide level precision. SOAP also shows very high contig level precision in the human dataset because it generates many very short candidates on this dataset. SOAP in total generates $153618$ predicted transcripts, but only $65562$ of them are longer than $100bps$. Excluding these exceptions, Bermuda has better precision than the other three assemblers (Trinity, IDBA and SOAP) at both contig and nucleotide levels. 
\subsection{Completeness distribution of predicted transcripts}
\begin{figure}[!Ht]
\centering
\includegraphics[width=0.46\textwidth]{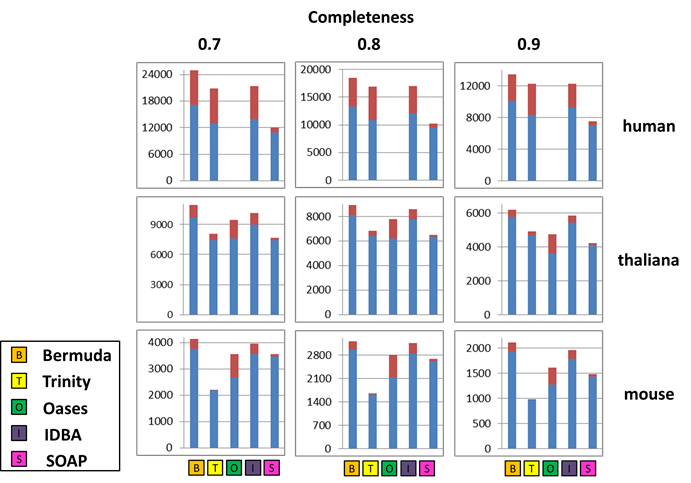}
\caption{Redundancy at different completeness levels. The nine figures show the comparison of redundancy of the five assemblers over human, thaliana and mouse data sets at completeness levels of $0.7$, $0.8$ and $0.9$, respectively.}
\label{Completeness}
\end{figure}
Figure \ref{Completeness} shows the redundancy of assembly results at three different completeness levels: 0.7, 0.8 and 0.9. In this figure, red color indicates the number of redundant occurrences, and the blue color indicates the total number of candidates excluding redundancy. As show in Fig 5, Bermuda outperforms the other assemblers in generating non-redundant candidates at each completeness level (marked by blue color). This also indicates that Bermuda outperforms the other assemblers in terms of recovering a real transcript to a better completeness level.
\subsection{Relationship with expression level} 
Table 2 shows the number of reconstructed real transcripts (with completeness level $0.8$) by the five assemblers with respect to expression levels. As described in $3.2$, we divide all the cDNAs into five groups according to their expression levels: $0-20\%$, $20-40\%$, $40-60\%$, $60-80\%$ and $80-100\%$. As shown in Table 2, Bermuda outperforms all the other assemblers at each expression level on all the three data sets. This shows that Bermuda can deal with unevenly distributed RNA-Seq data pretty well. 

\subsection{Computational Resources}
Table 3 shows the running time and peak memory usage of the four assemblers, tested on three datasets SRR364830 (human), SRR391051 (thaliana) and SRR404355 (mouse). All the assemblers are run with 1, 2 and 4 cores, respectively, on the same Linux compute node with 126G RAM. As shown in Table 3, Bermuda is slightly slower than SOAP, but much faster than the others. Bermuda consumes moderate amount of memory while Trinity and Oases consume too much memory, which may not be available to many users. The memory consumption by SOAP and IDBA also increase fast along with data size. 

\section{Discussion}
This paper has presented a new method Bermuda for de novo transcript assembly. This method is designed to handle transcripts with uneven and low expression levels as well as biased read coverage. Experimental results show that our method not only outperforms a few popular de novo assemblers in terms of the number of recovered transcripts, but also has favorable running time and moderate memory usage. \par
We will extend our work along two lines. One is to design a better method for isoform de-convolution by better utilizing the quantification information (i.e., the observed number of reads mapping to each node of the splicing graph) and formulating it as a regularized network flow problem. We will study an efficient algorithm that can decompose the flow of one splicing graph into several paths, each corresponding to one transcript, subject to the sparsity constraint. Currently our method is mainly designed for short read assembly, we will develop a new method that can combine both short and long reads \cite{koren2012hybrid} to further increase assembly accuracy. 

\section{Acknowledgments}
This work was financially supported by the National Science Foundation CAREER award (to JX) and the Alfred P. Sloan Fel-lowship (to JX). \\
Conflict of interest statement. None declared.

\bibliographystyle{abbrv}
\bibliography{bermuda}  

\balancecolumns
\end{document}